\documentclass[draft]{agujournal2018}
\usepackage{apacite}
\usepackage{url} 
\usepackage{lineno}
\usepackage{amssymb}
\draftfalse

\journalname{Space Physics and Aeronomy: Magnetospheres}

\begin{document}
\let\citep\shortcite
\let\citet\shortciteA

%
%


\title{Mini-magnetospheres and Moon-magnetosphere interactions:
  Overview Moon-magnetosphere Interactions}

%
%




\authors{Joachim Saur}

\affiliation{1}{Institut f\"ur Geophysik und Meteorologie,
  Universit\"at zu K\"oln, Cologne, Germany}





\correspondingauthor{Joachim Saur}{saur@geo.uni-koeln.de}




\begin{keypoints}
\item Review of moon-magnetosphere interaction (MMI) in the solar system
\item MMI is generally sub-Alfv\'enic and generates Alfv\'en wings
  with auroral footprints on planets
\item Observations of MMI can reveal atmospheres, plumes, ionospheres, subsurface oceans, and dynamos 
\end{keypoints}

%
%


\begin{abstract}
''Moon-magnetosphere interaction'' stands for the interaction of
magnetospheric plasma with an orbiting moon.
Observations and modeling of moon-magnetosphere interaction is 
a highly interesting
 area of space physics because it helps to better understand
the basic physics of plasma flows in the universe and it provides
geophysical information about the interior of the moons.  
Moon-magnetosphere interaction is caused by the flow of 
magnetospheric plasma relative to the orbital motions of the
moons. The relative velocity is usually slower than the Alfv\'en
velocity of the plasma around the moons. Thus the interaction
generally forms Alfv\'en wings instead of bow shocks in front of the moons.
The  local interaction, i.e., the
interaction within several moon radii, is controlled by properties of the atmospheres,
ionospheres, surfaces, nearby dust-populations, the interiors of the
moons as well as the properties of
the magnetospheric plasma around the moons. The far-field interaction,
i.e., the interaction further
away than a few moon radii, is dominated by the
magnetospheric plasma and the fields,
but it still carries information about the properties of the moons. In
this chapter we review the basic physics of moon-magnetosphere
interaction. We 
also give a short tour through the solar system
highlighting the important findings at the major moons.
\end{abstract}

%
%

%


%
%
%
%

\section{Introduction}
\label{s:intro}
The outer planets of the solar system all
possess large magnetospheres and harbor many natural satellites (for
simplicity also
called moons) within their magnetospheres.
Thus the phenomenon 
of moon-magnetosphere
interaction commonly occurs at the moons of Jupiter, Saturn,
Uranus and Neptune and only rarely at Earth,
when the Earth moon passes through the tail of its magnetosphere. 

Moon-magnetosphere interaction is a sub-class of  the interaction of a 
moving magnetized plasma with a celestial body. In terms of the very basic
physics, there is no fundamental difference between a moon interacting with
moving magnetospheric plasma or a planet interacting with the
solar or a stellar wind or even an artificial satellite interacting
with its plasma surroundings. In the solar system, however, under usual
circumstances the relative velocities between the solar wind and the
planets 
are 
larger than any of the three magneto-hydrodynamic (MHD)
wave modes and bow shocks form.
This is different in moon-magnetosphere interaction, where
 the relative velocities of the magnetospheric plasmas and
the moons embedded within them are usually smaller than the wave velocity of
the fast and Alfv\'en wave modes. Thus no bow shocks form around
the moons. The interaction is sub-Alfv\'enic and so called Alfv\'en
wings are being generated, which electromagnetically couple the moons
and the planets.

\subsection{Motivation}
Why do we care to understand moon-magnetosphere interaction? Studying
this interaction is important for two reasons: (a) The flow of plasma
around an obstacle is a fundamental physical process of
plasma, space and astrophysics and thus of
basic interest. (b) The moons and their properties modify the
plasma and magnetic field environment around the moons. Thus
observations of the space environment around the moons through space
probes and telescopes provide information about the properties of the
moons which are often not accessible otherwise. Such
plasma and magnetic field observation led, for example, 
to the discovery of plumes on Enceladus, and subsurface
water oceans within Europa and Ganymede.

\subsection{A short history}

The progress of understanding moon-magnetosphere interaction is
strongly tied to spacecraft missions to the outer solar system but
it also has important contributions from observations with telescopes.  
Moon-magnetosphere interaction was indeed first discovered
remotely in  Jupiter's radio emission, which contains a 
contribution that originates from Io's interaction with Jupiter's
magnetosphere \citep{bigg64}. Io is historically the body with the best studied 
moon-magnetosphere interaction because it is the most powerful
one in the solar system. The focus for decades remained with Io, where
Pioneer 10 in the 1970s detected Io's ionosphere \citep{klio75}, an important
ingredient in its moon-magnetosphere interaction. The Voyager 1 and
Voyager 2 spacecraft found that Io orbits within a dense plasma and
neutral torus generated by mass loss from the moon \citep{broa79,brid79}. Voyager 1 made the
first in-situ detection of moon-magnetosphere interaction when it
passed south of Io \citep{acun81} and provided evidence for Io's
Alfv\'en wings \citep{neub80}.  

Observational evidence for moon-magnetosphere interaction at the other
three Galilean satellites Europa, Ganymede and Callisto came with the
Galileo mission to the Jupiter system beginning in 1995 with several close encounters
at each of the moons \citep{kive04}. The Cassini spacecraft played the
analogues role in the Saturn system starting 2004. It provided
first observations of moon-magnetosphere interaction at the many inner
icy moons of Saturn and visited its largest satellite Titan more than a
100 times. 

Another milestone in remote sensing of the moon-magnetosphere
interaction came through the
discovery of so called auroral footprints of the moons. Again
beginning with Io, observations in
the infra-red \citep{conn93}, followed by the UV, and the visible
wavelength range, successively led to  the detection of the footprints of Europa, Ganymede, Callisto
and Enceladus so far \citep{clar02,pryo11,bhat18}. 

These observations sparked theoretical progress of the plasma-physical
processes generated by the moons early on.  
\citet{gold69} developed for Io's interaction the so called unipolar
inductor model, i.e., a steady-state electric current loop model. 
After the discovery of the dense Io torus, \citet{neub80},
\citet{goer80} and \citet{sout80} however realized that basics of the
interaction are best captured by considering the MHD waves generated
by Io, which led to the family of Alfv\'en wing
models. Further theoretical progress followed over the years where in
addition to analytical descriptions, numerical simulations of the
interaction nowadays play a very prominent role.

In this chapter we review basic principles of moon-magnetosphere
interaction (Section \ref{s:setup}). We classify the large variety of possible effects in
subgroups (Section \ref{sec:classification}). Then we discuss the plasma physics of the interaction close to
the moons and far away from them (Section \ref{sec:physics}). Finally we take a short tour to the
moons in the solar system where moon-magnetosphere interaction occurs
and 
we briefly look at extrasolar systems (Section \ref{sec:tour}). We end with a discussion of
outstanding issues (Section \ref{sec:questions}). 
This chapter focuses primarily on large scale, i.e., MHD effects
  of the interaction. Due to the page limit we discuss kinetic effects
  to a smaller extent and we omit moon-radiation
  belt interactions. The latter have led among other things to the
  discovery of a few small moons and can be considered a special
  case of the high energy tail of the plasma-moon interaction.

\section{Basic setup}
\label{s:setup}

\begin{figure}[h]
 \centering
 \includegraphics[width=10cm]{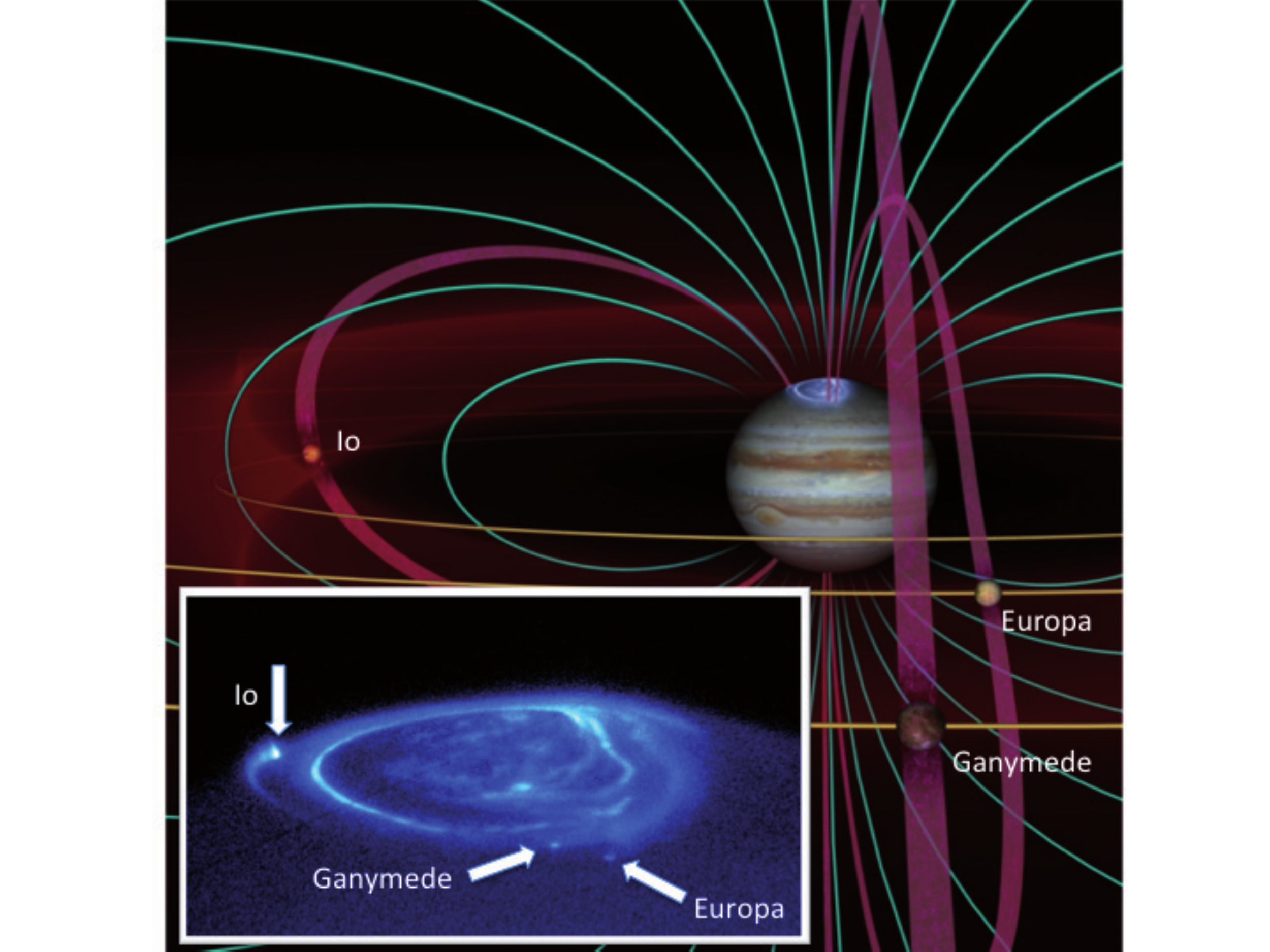}
 \caption{Set up of moon-magnetosphere interaction for the case of
   Jupiter's Galilean moons. The blue inlet is
a zoom into Jupiter’s polar region and displays Jupiter's
main aurora and the auroral footprints of the moons Io,
Europa, and Ganymede. Turquoise lines represent
Jupiter's magnetic field lines, and the orange lines show
the orbits of the moons. The pink flux tubes display field
lines connecting the moons with Jupiter.
Electric currents and energy generated by the
moon interaction is being transported  along these flux
tubes. Where the flux tubes intersect with
Jupiter's atmosphere, auroral emission is excited. The
plasma generated by Io produces a torus around Jupiter
(shown in orange/red). The radial transport of this
plasma generates magnetic stresses and electric currents
(not shown), which couple to Jupiter and generate its
main auroral oval.
(Image Credit: John Spencer and John
   Clarke).
} 
 \label{fig:Jupiter}
  \end{figure}

In Figure \ref{fig:Jupiter} we show the basic setup of
moon-magnetosphere interaction, where the Jupiter system is used as an
example.
The moons are embedded within the closed magnetospheric field lines of the
central planet and revolve about the planets with Keplerian
velocities (see orange lines
in Figure \ref{fig:Jupiter}). 
Their orbital periods are on the order of days to tens of days. 
The magnetospheric plasmas are frozen into the
magnetospheric fields and partly corotate with the planets, which have rotation
periods on the order of tens of hours. Due to an incomplete coupling
to the planets' ionosphere, the magnetospheric plasma is 
only partly corotating with the planet, which is referred to as
''sub-corotation''. The resulting azimuthal velocities of the subcorotating
plasma at the orbital distances of the moon is however still much larger than
the orbital velocities of the moons, which results in a relative velocity
between the magnetospheric plasma and the moons. In other words, the
moons are constantly overtaken by the magnetospheric plasma. 

Therefore the  moons are obstacles to the flow of magnetospheric plasma and
thus perturb the plasma and magnetic field environment around them. 
This interaction  generates waves, which propagate away from
the moons. The most important wave is  the Alfv\'en mode, whose group
velocity in the rest frame of the plasma travels strictly parallel and anti-parallel to the ambient
magnetic field. In the rest frame of the moon, this leads to standing
Alfv\'en waves, called Alfv\'en wings. Where these wings intersect
with the planets' atmosphere/ionosphere auroral emission is generated,
which is referred to as footprints of the moons (see Figure \ref{fig:Jupiter}).

\section{Classification of interactions}
\label{sec:classification}
Now we study how the various interactions at the moons in the solar
system can be classified. Starting from the general case of the
interaction of planetary bodies within any type of moving magnetized plasma, we 
subdivide the interaction into several subclasses. Two basic 
criterion
 to
categorize the interaction are (a)  the properties of the external
plasma which the moon is exposed to and (b) the properties of the
moon. We will systematically discuss these properties in the
next two subsections. An additional 
criterion
is the nature of the
coupling between the moon and the host planet, which
we subsequently present in Section \ref{ssec:prop_coupling}.

\subsection{Properties of the magnetospheric plasma}
\label{ssec:prop_plasma}

In most cases moon-magnetosphere interaction can be described within
the magnetohydrodynamic (MHD) framework. MHD is applicable if the ion gyroradius of the
plasma is significantly smaller than the spatial scales of the moons
and if the ion gyroperiod is significantly smaller than typical
time scales, e.g., the convection time of the plasma past the
moon. In nearly all the cases of moon-magnetosphere interaction, this
is a very good assumption. We will discuss some exceptions and
modification of moon-magnetosphere interaction through non-MHD effects
in Section \ref{sec:tour}.

The MHD wave modes, i.e., the shear Alfv\'en mode, the
magneto-acoustic modes 
and the convecting entropy mode play the most
important role in the interaction. 
%
Basic properties of these modes are
given by the Alfv\'en velocity $v_A$ and the sound velocity $c_s$, which
lead to the Alfv\'en Mach number 
\begin{eqnarray}
M_A=\frac{v_0}{v_A} \hspace{1cm} \mbox{with} \hspace{0.5cm}
  v_A=B/\sqrt{\mu_0 \rho} \;,
\label{e:vA}
\end{eqnarray}
the sonic Mach number
\begin{eqnarray}
M_S=\frac{v_0}{c_s} \hspace{1cm} \mbox{with} \hspace{0.5cm}
  c_s=\sqrt{\gamma p/ \rho} \;,
\end{eqnarray}
and the fast Mach number
\begin{eqnarray}
M_f=\frac{v_0}{\sqrt{c_s^2+v_A^2}}=\frac{1}{\sqrt{\frac{1}{M_s^2}+\frac{1}{M_A^2}}}
\end{eqnarray}
with the relative velocity between the magnetospheric plasma and the
moon $v_0$, the magnetic field strength $B$, the plasma mass density $\rho$,
the total plasma pressure $p$,
i.e. the sum of the ion and electron thermal pressure and the
adiabatic exponent $\gamma = 5/3$. 
The definition of the Mach numbers directly imply $M_f \le M_A$ and $M_f
\le M_S$.
The Alfv\'en and sound speeds also constrain the plasma $\beta$, i.e., the ratio between
the thermal and the magnetic pressure given by
\begin{eqnarray}
\beta=\frac{p}{B^2 / 2 \mu_0}=\frac{2}{\gamma}
  \left(\frac{M_A}{M_s}\right)^2 \;.
\end{eqnarray}

The most important classification criterion is whether the fast Mach
number $M_f \gtrsim 1$. 
Then waves excited by the moon cannot
propagate upstream because the flow velocity is larger than the
largest wave speed. In this case a bow shock forms. In nearly all the
cases of moons in planetary magnetospheres the Alfv\'en Mach number
obeys $M_A< 1$, thus $M_f<1$ as well, and no bow shock in front of the
moon develops. This type of interaction is referred to as
sub-Alfv\'enic interaction. The most prominent exception can be Titan
and possibly Callisto.

Another important criterion is the plasma beta. In case of
sub-Alfv\'enic interaction, low plasma beta $\beta \ll 1$ implies the interaction is
controlled by the magnetic energy of the magnetospheric plasma. This
generates a ''stiff'' magnetic field environment. In case of large
plasma beta $\beta \gg 1$, the thermal energy of the plasma dominates
the interaction. In this case, the magnetic field lines are strongly draped
around the obstacle because the thermal pressure dominates the flow
around the obstacles and takes the magnetic field via the
frozen-in-field theorem with it.

\subsection{Properties of the moons (atmosphere, ionosphere, dynamo,
  oceans)}
\label{ssec:prop_moons}

The various properties of the moons influence their space environment
as well. The primary root cause of the interaction is that the
moons act as obstacles and modify the flow of plasma around them through mechanical or
electromagnetic forces. 

\subsubsection{Mechanical Obstacles}
Mechanical obstacles are the solid surfaces of the moons, which absorb
the plasma on the upstream side and which generate wakes of void
plasma on the downstream side. The moons' atmospheres and exospheres
including dust are also mechanical obstacles because collisions with the plasma
and ionization cause a modification of the velocity and momentum of
the plasma. 
Collisions can pertain to elastic collisions or
 charge-exchange collisions. Ionization of neutrals leads to fresh ions
 and electrons, which are subsequently reaccelerated by the ambient 
motional electric fields. This processes is refereed to as
pickup
\citep{vasy16}. 
The effects of collisions and ionization can be formally combined to
'effective collision frequencies' used in the electron and ion 
equations
\citep{neub98}.
Collisions and pickup both cause the electrical
conductivities in the moons ionospheres.

\subsubsection{Electromagnetic Obstacles}
Moons with an intrinsic magnetic field interact electromagnetically with the
surrounding plasma. The intrinsic magnetic fields can be due to an
internal dynamo field, such as within Ganymede
\citep{kive96b}. Other possibilities are induced magnetic fields in the
interior. Time-variable external magnetic fields generate secondary
magnetic fields within 
electrical
conductive layers such as
internal saline
water oceans. Those secondary fields modify the
magnetic field and plasma flow around the moons
\citep{khur98,neub98,schi07}.

\subsection{Unipolar inductor vs Alfv\'en wing model}
\label{ssec:prop_coupling}
Moon-magnetosphere interactions generate Alfv\'en waves, which travel
along the host planets moving magnetospheric field lines and eventually
hit the planet. 
Where the Alfv\'en waves intersect with the
atmosphere and ionosphere of the planet auroral emission is being
observed \citep{conn93}. The Alfv\'en waves are partially reflected at the
planet's ionosphere and travel back towards the moon. When the
reflected wave misses the moon because the magnetospheric plasma has
traveled far enough downstream, then no feedback coupling between
the moon and planet is possible. This case is called the pure Alfv\'en
wing model. It occurs when the wave travel time from the moon to the
planet and back $2 \tau_\mathrm{wave}$ is larger than the convection time of the
plasma past the moon $\tau_\mathrm{conv}$ \citep{neub80,neub98}. If the wave
travel time is much shorter than the convection time, then a force
balanced situation with only field-aligned currents between the moon
and the planet is reached, which is referred to as the
'unipolar inductor case' \citep{gold69}. The situation where the waves
can only partially couple back, can be referred to as the mixed Alfv\'en
wing case \citep{neub98}.

\section{Physics of the interaction}
\label{sec:physics}

The physics and the effects of moon-magnetosphere interaction can be
conveniently divided into the physics of the local interaction, i.e.,
within several radii of the moon. In this region  the
detailed nature of the obstacle plays an important role in controlling
the space environment around the moons. The regions further away is
often referred to as far-field. It includes for example the Alfv\'en
wings, the auroral footprints and their tails.

\subsection{Physics of the local interaction (i.e., within several moon
  radii)}
\label{ss:local}

The root cause of the interaction occurs in the local
interaction. Mechanical or electromagnetic forces slow and modify the
magnetospheric plasma flow. The locally modified flow generates magnetic
field perturbations and also leads to MHD and other wave modes. 
The local
interaction is generally very complex because the different wave modes
interact non-linearly. 
In the local interaction the
detailed aeronomic processes that occur in the atmospheres and exospheres need to be
considered as well.

In the following we mostly discuss cases of moon-magnetosphere interaction where
at the surface of the moon the magnetospheric field of the central
planet is larger than the magnitude of the internal magnetic field. In this
way, no closed magnetic field lines region around the moons
develop. The only known moon with closed field lines is Ganymede
\citep{kive96b,neub98,jia08,duli14}. Ganymede's space plasma
environment is reviewed in a separate chapter by X. Jia et al. 

\subsubsection{Plasma flow in the atmosphere and magnetic draping}
\label{sssec:draping}

\begin{figure}[h]
 \centering
%
 \includegraphics[width=12cm]{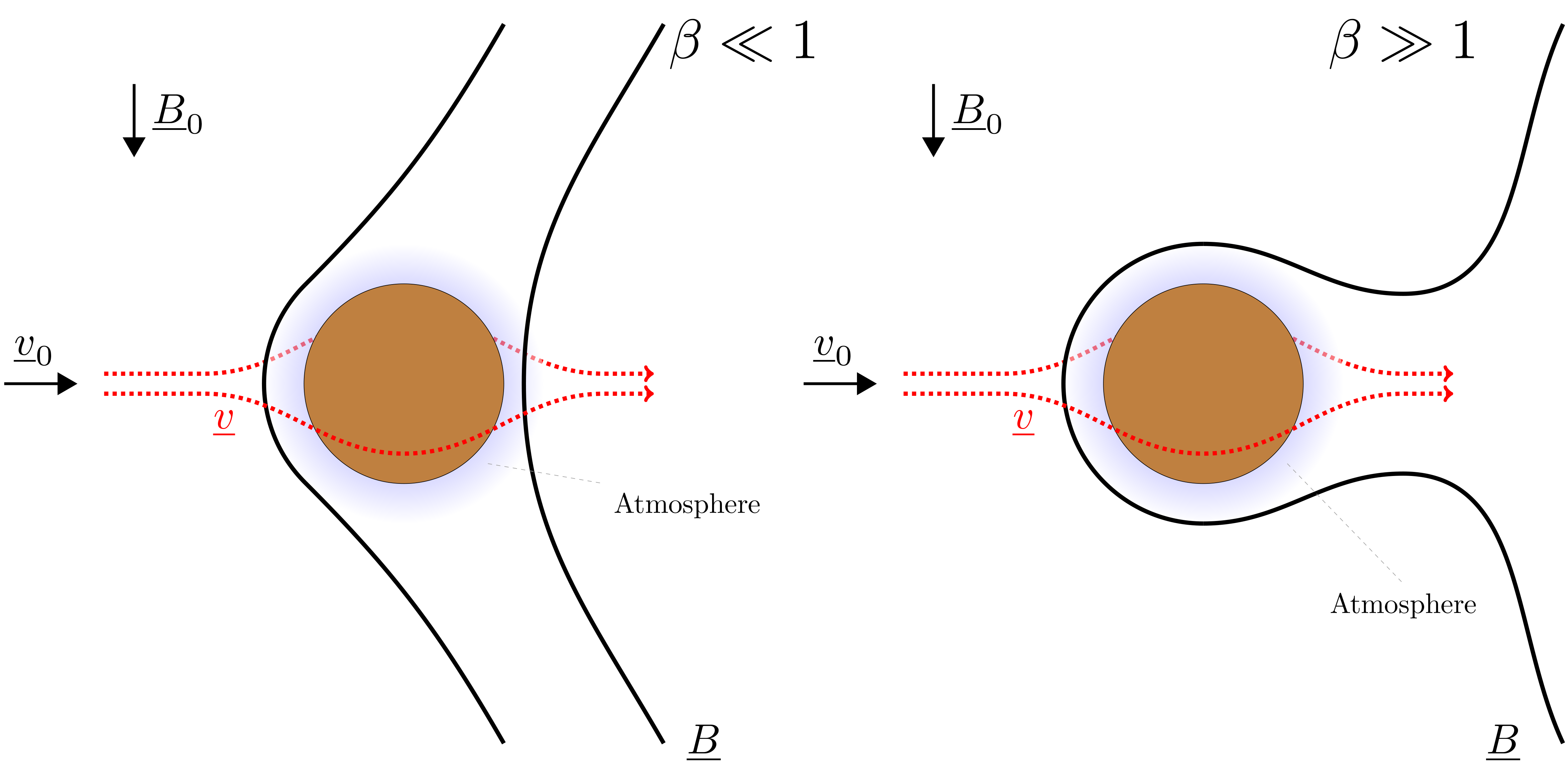}
 \caption{Sketch of local plasma interaction in the sub-Alfv\'enic
   case for (a) low plasma beta $\beta \ll 1$ and  (b) high plasma beta $\beta >1 $.} 
 \label{fig:local}
  \end{figure}

In Figure \ref{fig:local}
we sketch the local interaction in the
sub-Alfv\'enic case for (a) low plasma  $\beta \ll 1$, e.g., Io's
environment, and  (b) for high plasma  $\beta >1 $, e.g., Titan's
environment. 
In both cases charge-exchange and collisions between magnetospheric
ions and atmospheric neutral particles strongly slow the initially fast moving plasma.
Pickup, i.e., ionization of the
neutrals, which move with very slow velocities in the rest frame of the
moons, adds new slow-moving plasma to the magnetospheric plasma and thus
slows the total plasma as well. 

The slowed plasma flow causes a change in the surrounding magnetic
field. The plasma outside of a  moon's atmosphere obeys the
frozen-in-field theorem. The flow outside of the atmosphere is
generally much faster compared to the flow in the atmosphere which
leads to draped magnetic field lines around the moons. On the flanks
of the moons, the velocity can be accelerated to velocities
up to twice the upstream
velocity $v_0$ \citep{saur04}. The draping is
particularly strong in cases of large plasma beta and much weaker for low
plasma beta (see Figure \ref{fig:local}). The reason is that in the
low beta case, the energy density in the magnetic field dominates the
thermal energy density. This is often the case for moons which are
closer to the planets. For low plasma betas, the strong background magnetic field
acts as a stiff magnetic field, which controls the topology of the
field. It results in larger magnetic tension forces compared to the pressure
gradient, which prevents the draping. Next to the magnetic field
direction the field magnitude is also modified in the interaction. On
the upstream side the magnetic field strength is enhanced due to
compressional effects, which is referred to as magnetic pileup, while
on the downstream side the field magnitude is decreased compared to the
unperturbed magnetic field.

\subsubsection{Aeronomy and formation of ionospheres}
Even though the main engine of moon-magnetosphere interaction is the
momentum exchange between the magnetospheric plasma and the moon, the
detailed processes in the moons' atmospheres and ionospheres are often
controlled by complex aeronomic processes. The two main ionization
processes in the moons' atmospheres are electron impact ionization and
photo-ionization. Electron impact ionization can occur through the
thermal and non-thermal electrons in the planets' magnetospheres. 
Energetic electrons produced by the interactions can additionally
contribute to ionization.
This can be electron beams  generated within the
Alfv\'en wings, which return to the moon and
generate a feedback loop within the interaction
\citep{will96,jaco10}. At Ganymede the interaction drives additional
sources of energetic electrons in form of reconnection and wave
particle interaction
 within and near the closed field line region of Ganymede's
magnetosphere \citep{will97,evia01}. 
The ionospheres are often not in
chemical equilibrium, i.e., ionization is not balanced by recombination,
but additionally controlled by convection of ionospheric plasma. Convection
thus plays an important role in shaping the spatial structures of the moons
ionospheres.
Electron impact ionization rate depends on the neutral
and electron density, but also highly non-linearly on the electron
temperature or its distribution function \citep{rees89}.

\subsubsection{Interaction strength and ionospheric Pedersen and Hall
  conductances}
\label{sssec:iono}

The presence of a magnetic field makes the electrical conductivity of a
plasma anisotropic. Within the ionospheres of the moons, the
conductivity tensor is conveniently separated in Birkeland
conductivities $\sigma_B$ (parallel to the magnetic field), in Pedersen
conductivities $\sigma_P$ (perpendicular to the magnetic field, but still
along the electric field), and the Hall conductivities $\sigma_H$
(perpendicular to the magnetic and electric field). Except for the case
of Titan, the Birkeland conductivities can be considered nearly infinite
within the moon's ionospheres \citep{neub98} which leads to negligible
field aligned electric fields in the ionospheres. Therefore the
Pedersen and Hall conductivities can be integrated along the field
lines to obtain height-integrated conductivities, referred to as
Pedersen and Hall conductances, $\Sigma_P$ and $\Sigma_H$,
respectively. The values and the spatial structure of these
conductances strongly constrain the ion and electron flow in
the moons' ionospheres.

The conductances can be used to characterize the strength of the
moon-magnetosphere interaction. The interaction strength
$\bar{\alpha}$ quantifies how strongly the magnetospheric flow is
reduced within the moons' ionospheres \citep{saur13}. It is defined by
\begin{eqnarray}
\bar{\alpha} \equiv\frac{\delta v}{v_0}  \approx \frac{1}{M_A} \frac{\delta B}{B_0}
\label{e:alpha_bar}
\end{eqnarray}
with $\delta v = |v - v_0|$ and $\delta B = |B - B_0|$ the amplitudes of the velocity and the
magnetic field perturbations in the Alfv\'en wings or approximately
in the vicinity of the moons.
 The interaction strength
varies between zero and its maximum value one, when the plasma 
flow comes to a halt near the moons, i.e., $v=0$. In case of an
Alfv\'enic  far-field interaction,
the interaction
strength can be approximated as $\bar{\alpha}=\Sigma_P/(\Sigma_P + 2 \Sigma_A)$ 
or by the ratios of the moon's ionospheric conductances to those of
the planet in case of the unipolar inductor far-field interaction \citep{gold69,saur04}.

The ionospheric Hall effect  breaks the interaction symmetry and
rotates the flow and the magnetic field. The ionospheric electric
field and the electron flow are rotated by the angle
$\Theta_\mathrm{twist}=-\Sigma_H/(\Sigma_P+ 2 \Sigma_A)$ \citep{saur99a}. The
magnetic field perturbations in the Alfv\'en wings are rotated by the
same angle. At the presence of a significant dust component in the
plasma, a fraction of the electrons might be attached to the mostly
immobile dust grains. Accordingly ions are not the most massive charge
carrier in the plasma anymore. This can generate an anti-Hall effect with
reversed sign of the Hall conductances and thus
reversed directions of the electron flow and fields
 as observed within the plumes
of Enceladus \citep{simo11,krie11}.

\subsubsection{Overall energy fluxes}
It is interesting to look at the overall partioning of the energy
fluxes in moon magnetosphere interactions. For simplicity we only
consider the low
Alfv\'en Mach number
 $M_A \ll 1$ and the low plasma beta case $\beta \ll 1$ 
under the assumption that
the unperturbed plasma velocity  $v_0$ and magnetic field $B_0$ are
perpendicular. The largest energy flux, which the moon's
atmosphere/ionosphere system is exposed to,
is the Poynting flux of the magnetospheric plasma, i.e., the bodily
transport of magnetic enthalpy.
Assuming the ionospheric obstacle
has a radius of $R$, then the Poynting flux of the upstream flow is $S_{in} = \pi 
R^2 E_0 B_0/\mu_0$ with the motional electric field $E_0=v_0 B_0$. 
The energy dissipated as Joule
heating in the moons'
ionospheres can be approximated by $P= 4 \pi \; \bar{\alpha} (1
-\bar{\alpha} ) M_A  R^2 
 E_0 B_0  / \mu_0$ \citep{neub80}. The Joule
dissipation $P$ is maximum for intermediate interaction strength $\bar{\alpha}=1/2$. 
The total energy radiated away within both Alfv\'en wings towards the planet is
$ S_{A} = 4 \pi \; \bar{\alpha}^2 M_A  R^2  E_0 B_0 /\mu_0$ \citep{saur13}. 
The ratio $S_A/P= \bar{\alpha}/(1- \bar{\alpha})$ depends only on the
interaction strength $\bar{\alpha}$. 
The Alfv\'enic Poynting $S_A$ flux is larger than the ionospheric Joule
dissipation for $\bar{\alpha} > 1/2 $  and $S_A $ assumes its maximum 
if the interaction strength is maximum, i.e.
$\bar{\alpha}=1$.  
How much of the Alfv\'enic Poynting flux in the wings 
directly enters and heats the planets' ionospheres strongly depends
on the nature of the far-field coupling \citep{gold69,neub98}.

\subsubsection{Induction in the interiors: Diagnosing saline oceans}
Time-variable magnetic fields outside of the moons induce electric
fields according to Faraday's law of induction. Any electrically
conductive layers will then generate electric currents which produce
secondary magnetic, i.e., induced magnetic field. In case of
Jupiter, Uranus and Neptune, the  magnetospheric fields in the rest frame of the
moons are quasi-periodic due to the tilt of the magnetic moments of the
planets with respect to their spin axis.  Observations of induced
magnetic fields outside of the moon are thus diagnostic of
electrical conductive layers
\citep{khur98,kive00,zimm00}. This method is a very powerful tool to detect
currently liquid, electrically conducting oceans under the icy crust of some of the moons, in
particular Europa and Ganymede. The method is based on the fact that the
electrically conductivity of a mantel composed of pure ice and
rock compared to a saline ocean as expected for these icy satellites is several orders of
magnitudes smaller and would not produce observable magnetic fields \citep{khur98}.

For a quantitative analysis of the induction effects, both the inducing and the induced fields need
to be known. Therefore in general two satellites which measure both
fields separately are necessary. In case of quasi-periodic
fields, the inducing fields from magnetospheric measurements (often combined
with theoretical models) can be used such that single flybys at the
moons are sufficient. However attention still need to be paid to
separate magnetic fields generated by internal induction from other
sources of magnetic field perturbations, e.g., due to
the plasma interaction with the atmosphere \citep{schi07,schi08}.
 
In case the inducing magnetospheric fields at the moon can be
considered
spatially  constant over the size of the moon, which is usually a
very good assumption, and under the assumption of a 
radially symmetric conductivity distribution, 
the induced field is a dipole magnetic
field outside of the moon \citep{zimm00}. Thus observations of a
dipole field perturbation with the appropriate phase with respect to the
inducing field allows to derive constraints on 
electrically conductive layers.  

The induction sounding technique is
non-unique. Observations of one single inducing frequency does not
allow to separate thickness and conductivity of the electrically
conductive layer, but only approximately constrain
the product of both \citep{zimm00,seuf11}. The analysis of
magnetic field measurements obtained during multiple flybys or from an
orbiter around the moons is thus a powerful tool to probe various
inducing frequencies and thus to disentangle various subsurface
ocean properties 
{\citep{khur02,seuf11}}
 (as 
planned 
for the Europa Clipper and the Juice Missions).

\subsubsection{Induction in the ionospheres}

In addition to induction in the interior of the moons,
time-variable magnetic fields can also generate induced fields in the
electrically conductive ionospheres of the moons. The 
conductivity in an ionosphere is anisotropic (see Section
\ref{sssec:iono}). In saline oceans
or in electronically conductive metals and rocks the conductivity is 
however approximately isotropic. Recently it was shown
that induction in an ionosphere 
is controlled by an effective conductance $\Sigma_\mathrm{eff}= \Sigma_P +
\Sigma_H^2/(\Sigma_P+\Sigma_A)$ \citep{hart17a}. In case of large Hall
conductances $\Sigma_H$ and small Alfv\'en conductances,  
a significant enhancement effect occurs compared to the individual
Hall and Pedersen conductances. The underlying process is similar to
the Cowling channel effects known from the Earth's ionosphere
\citep{cowl32,baum12}. Because the ionospheric conductivities depend on
the neutral and the ion densities and are inversely-dependent on
the magnetic field strength, ionospheres with large densities or moons
at large orbital distances within smaller magnetospheric fields have
large effective conductances and will thus generate sufficiently strong induced
fields. In the case of Callisto, the effective ionospheric
conductances can be as large as the ocean conductances
\citep{hart17a}. 
Thus observations of induced fields do not
necessarily imply the existence of a subsurface oceans. 
Thus observations of induced fields at distances significantly 
above the ionosphere
  do not necessarily imply the existence of a subsurface
  ocean. Observations, which better constrain the spatial distribution
  of Callisto's ionospheric and neutral
  densities are necessary to better quantify the ionospheric
  contribution to induction, particularly on the night side. If
  possible, flybys below the peak of the ionospheric densities would
  help disentangle induction effects in ionospheric and internal layers.

\subsubsection{Wakes, upstream-down stream asymmetries}

Moon-magnetosphere interactions possess a pronounced
upstream-downstream asymmetry (see also section \ref{sssec:draping}). 
Compared to hydrodynamic flows around solid objects, the flow of
plasma onto moons with very dilute atmospheres leads to absorption of
plasma on the surfaces of the upstream side. This effect causes 
the formation of
a wake of void plasma on the downstream
side. The void plasma wake will be partially filled by plasma driven by
thermal pressure. In case of low Alfv\'en Mach number $M_A$ and small plasma
beta, the plasma will move along the magnetic field lines at a rate
increasing with the sonic Mach number $M_S $ \citep{neub98}. The
rarefied plasma in the wake leads to a reduced thermal pressure
compared to plasma on streamlines without absorbed plasma. The resultant
pressure gradients into the wakes are partially compensated by
enhanced magnetic pressure associated with diamagnetic currents. The
wake effects will thus
lead to enhanced magnetic field
magnitudes in the wakes. These perturbations have been observed and
modeled initially at the Earth's moon \citep{whan68} and subsequently 
studied for
example also at the nearly inert moons Tethys and Rhea
\citep{khur08,simo09,simo12}. In the case of Io, it can explain the
double-peaked magnetic field structure in its wake \citep{saur99a}. 

When strong mass-loading in the moons' atmospheres or exospheres occurs, 
in particular on the flanks, the new plasma is transported
downstream and produces  another part of
the  wake with enhanced density outside of the rarefied wake. 
This density will generate a density wave along
the magnetic field lines \citep{schi08}.

\subsubsection{North-South asymmetries}

 Asymmetries in the moons' neutral gas environment will generate
 asymmetries in the moon-magnetospheric interaction. Let us assume for
 simplicity that the background magnetic field is in the north-south
 direction. An atmospheric asymmetry between the north and the south,
 for instance,
 due to the plumes on Enceladus, will drive different
 ionospheric electric  currents in the north and the south. With the
 exception of Titan with its dense atmosphere, field lines which pass the
 moons' atmospheres, but which do not penetrate the solid body of the
 moons, are 
equipotentials  
due to very large field-aligned 
 conductivity  $\sigma_B$ within the thin atmospheres of the moons
 \citep{neub98}. Because the solid surfaces of the moons are
 electrically insulating, magnetic field lines which penetrate the
 surface do not need to be equipotentials anymore. In case of an
 atmospheric north-south asymmetry, different amounts of electric
 currents are driven in both hemispheres which are continued along the
 magnetic field lines in the northern and southern magnetosphere,
 respectively. The asymmetric amount of electric current between both
 hemispheres is partially reduced by hemispheric currents, which
 connect both hemispheres \citep{saur07}. These hemisphere currents
 generate rotational magnetic field discontinuities on field lines
 tangent to the solid body of the moons. Such discontinuities have
 been identified in the magnetic field environment of Enceladus
 \citep{simo14}. The hemisphere currents and discontinuities are
 suited to search for plumes and atmospheric asymmetries, but they are
less pronounced in case a globally symmetric 
atmosphere is present  \citep{bloe16}.

\subsubsection{Non-MHD effects}
Most of the large scale features of moon-magnetosphere interactions can be
described by resistive single fluid MHD when the ionospheric Hall effect is
included. Several of the following multi-fluid or non-MHD effects can however be
important as well: 
(1) In case the ion population cannot be appropriately modeled with one
effective ion, multi-ion fluid effects need to be considered
\citep{paty04,ma06}. 
(2) In some cases, it is important to consider the electron physics
separately. For example, electron heat
conduction can maintain ionospheric electron
temperatures at a level such that electron impact ionization is the
dominant ionization source
\citep{saur98,back05,rubi15}.
(3) In case photoionization is the dominant ionization source, 
strongly  non-Gaussian electron distribution functions can arise,
which require a kinetic, i.e. Vlasow-based description, e.g., at Titan and
Callisto
\citep{vigr16,hart17}. 
(4)
Due to the decreasing planetary magnetic field
strength with distance from the planet, large gyroradii-effects can
play a role at some of the moons. When ions are picked up on the flanks of the moons where
the convection velocity is large, gyroradii as large as the moons can
occur. 
\citep{kive04,liuz15}. 
Prominent examples are Titan and Callisto. However, a large fraction of the
pickup ions at these moons have still small gyroradii because the
ionization occurs within the bulk atmosphere where the plasma flow is
significantly slowed
\citep{hart17a}. 
(5) For the total plasma pressure, the suprathermal ions can play an
  important role. At Jupiter's Galilean moons, the pressure of the
  suprathermal magnetospheric ions can be comparable or larger than the
  thermal ion pressure    
\citep{mauk04} 
(6) Reconnection and particle acceleration processes at Ganymede's
magnetosphere are strongly controlled by kinetic ion and electron physics
\citep{toth16}.

\subsection{Physics of the far-field interaction: Alfv\'en wings and footprints}
\label{ss:farfield}

The far-field interaction is the region starting several moon radii
away from the moons. It does not include the moons' ionospheres any
more. The far-field is strongly controlled by the MHD wave modes.
The transition from
the local interaction into the far-field is included in all standard
3D numerical models and continuation of electric current from the local
interaction into the Alfv\'en 
wings is usually included in the analytic models as well.

\subsubsection{Role of MHD waves}
Let us look at the far-field effects at various limits. For simplicity
we first  assume that the magnetospheric fields are spatially and temporally
constant and the interaction is sub-Alfv\'enic. In the limit of
negligible plasma pressure, i.e., the approximately zero
plasma-beta, the propagating wave modes are the fast mode and the
Alfv\'en mode. The fast mode is in this case isotropic in the plasma
rest frame. Therefore the energy  flux density of the fast mode decreases with
distance squared. Thus the fast mode amplitudes are negligible at
sufficiently large distances from the moons.  The Alfv\'en mode, in
contrast, has a fully anisotropic group velocity, which is 
strictly parallel or anti-parallel to the background magnetic
field in the rest frame of the unperturbed plasma.  
The Alfv\'en mode is therefore the most important mode in the
far-field because the wave energy density does not decrease along the
path of its group velocity. The Alfv\'en mode can be conveniently 
described in Els\"asser variables \citep{elsa50}, which are also the
characteristics of the equations. They are  given by
\begin{eqnarray}
\bf{z^\pm}=\bf{v} \pm \frac{\bf{B}}{\sqrt{\mu_0 \rho}} \; .
\end{eqnarray}
In case of a north south orientation of the magnetospheric field  at
the location of the moon (like at Jupiter or Saturn), the $\bf{z}^+$ variable characterizes the
southern Alfv\'en wing and $\bf{z}^-$ the northern wing. 
The wings are inclined with respect to the background magnetic field
by an angle $\tan^{-1}
M_A$ \citep{neub80}. The reason is that while the wave propagates
along $\bf B_0$ 
it is simultaneously convected
downstream by $ \bf v_0$. 
In case that
the north and south propagating waves do not intersect and the
background fields are constant,  an Els\"asser variable being constant
constitutes an exact 
solution of the non-linear incompressible MHD equations. This is an important fact
because it describes that non-intersecting Alfv\'en waves can propagate without
dispersion and dissipation towards the central planet. 

\begin{figure}[h]
 \centering
 \includegraphics[width=6cm]{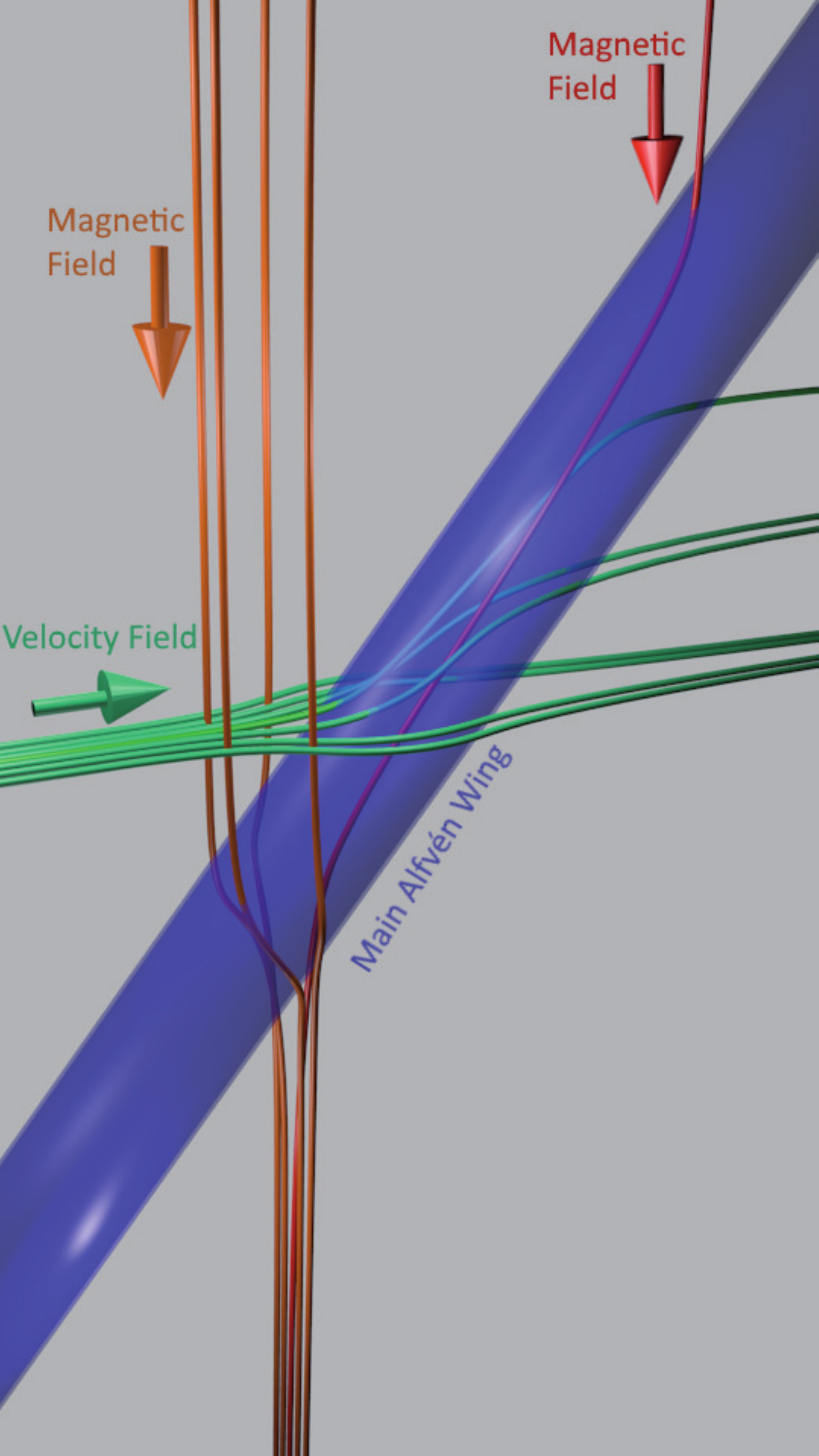}
 \caption{Magnetic field lines (orange and red) and velocity stream
   lines (green) within an
   Alfv\'en wing. Field lines in red pass through the main wing and
   the field lines in orange pass outside of the main wing.   
The purple tube characterizes the boundary of the main
    wing which corresponds to the size of the source, i.e., the
   moon. 
Inside the main wing the
   magnetic field lines are bent towards the direction of the tube,
   i.e., in downstream direction,   outside of the main wing on the
   flanks 
the magnetic field lines are bent in the opposite
   direction, i.e., upstream. 
} 
 \label{fig:far}
  \end{figure}

In Figure \ref{fig:far} 
we display the basic magnetic field and
velocity structure in an Alfv\'en wing. Velocity streamlines are shown
in green and magnetic field lines are in red and orange.
In the center of the main Alfv\'en wing (shown as purple tube) the
flow is slowed. This results 
from the slowed flow in the moons' atmospheres, which is 
caused by collisions of the plasma with the
  atmospheric neutrals and by pickup due to ionization of neutrals. The
  slowed plasma flow in the atmospheres is then propagated as
  Alfv\'enic velocity and associated magnetic field fluctuations into the Alfv\'en wing.
Within the Alfv\'en
wing the streamlines have a significant component parallel to the characteristic and thus
follow the wing over a certain distance. In case of maximum
interaction strength, i.e., $\bar{\alpha} =
1$, the flow within the main wing would only have a component parallel
to the tube and thus would stay within it.
Outside of the main wing,
the flow is directed around the region of slowed plasma with increased
velocities on the flanks to maintain incompressibility. The magnetic field lines
within the main wing (in red) are bend back compared to the background magnetic
field and follow the tube. In the case of $\bar{\alpha} = 1$, the magnetic field in
the main wing would be perfectly parallel to the wing and magnetic
field lines would stay within it as well. In all other cases
$\bar{\alpha} < 1$ , 
the magnetic field lines reside within the main wing for a certain distance and then turn
towards the unperturbed direction. 
Outside of the main wing on their flanks, the magnetic field lines (in orange) are bent in the
opposite direction compared to the direction of the bent within the
wing.  The resulting magnetic stresses are such that they are
balanced by acceleration or deceleration of the plasma. 

Around the moons and partly within the Alfv\'en wing the slow mode
plays an important role as well.
%
The propagation of the slow mode
strongly depends on the plasma temperature. If the sound speed is
much slower than the Alfv\'en velocity, i.e. $\beta \ll 1$, 
the slow mode propagates primarily
along the magnetic field with a slow mode wing inclined by $\tan^{-1}
M_S$. 
A sonic shock then propagates strictly along the magnetic field.
If the
sound speed is larger than the Alfv\'en velocity,
the sonic mode
propagates with the Alfv\'en velocity. The resultant wing contains
Alfv\'enic and slow mode features. It is fan-shaped in the plane
containing the background velocity and the magnetic field. The
structure is 
also referred to as delta wing and has an opening angle of approx 
$\tan^{-1}
M_A$ \citep{neub06}. 

\subsubsection{Kinetic effects in far-field}
A part of the far-field interaction is not controlled by ordinary
  MHD. Kinetic effects are responsible for the
 acceleration of particles which generate the footprints of the moons
in Jupiter's atmosphere through the electromagnetic wavelengths
spectrum 
\citep{conn93,pran96,clar96,pryo11} 
and for the radio-emission generated along the Alfv\'en wings
\citep{bigg64,zark98,zark18,loui17}. Due to the strong magnetic field
and the small plasma densities in this region, the
displacement current in Amp\'ere's law needs to be included
and for electron energies
larger than a few 100 keV the relativistic mass of
the electrons need to be considered. 

Theoretical ideas for the particle acceleration
mechanisms have been developed primarily for Io's Alfv\'en wings and
are mostly based on the filamentation of
the main wing \citep{chus05}. These filamented Alfv\'en waves are suggested
to turn into kinetic Alfv\'en waves and develop stochastic magnetic
field aligned electric fields, which can generate broad band electron
distributions \citep{hess10,hess11,hess13a,saur18a}.
 The
resultant electron beams can trigger the electron maser instability,
which causes the satellites' controlled 
radio emission
\citep{bigg64,zark98,zark01,zark07,zark18}. \citet{su02} investigated
electron acceleration downstream of the main wing with a 
steady-state Vlasov model by applying a potential
drop. Alternatively, \citet{crar97a} suggested that repeated Fermi-acceleration
can generate Io's electron beams.

\section{Short tour of moons in magnetospheres: Earth, Jupiter,
  Saturn, Uranus, Neptune, and Exoplanets}
\label{sec:tour}

Now we take a very short tour to the moons located within the
respective magnetospheres of their host planets.
Because of the very large number of moons in the solar system, we can
only cover a selected number of moons guided by scientific interest
and size. 
Thus we do not cover moons associated with planets or dwarf planets 
with at most only weak internal magnetic fields like Mars and Pluto.
We include
Earth's moon even though it is only a fraction of the time in the magnetosphere.
In case of Jupiter we cover the Galilean moons, in case of Saturn the
larger inner icy moons plus Titan, in
case of Uranus and Neptune we consider all moons with
diameters larger than 1000 km. These objects, their main properties and
associated references are listed in table
\ref{tab:moons}.
 \begin{sidewaystable}
  \caption{Overview of moon-magnetosphere interaction properties for
    selected solar system moons. Table includes: Radius, semi-major
    axis, magnetospheric field strength $B_0$ at location of moons,
    relative velocity $v_0$ between moons and magnetospheric plasma,
    magnetospheric plasma
    density $n_p$ at location of moon, Alfv\'en Mach number $M_A$, sonic
    Mach number $M_S$, plasma $\beta$ and interaction strength
    $\bar{\alpha}$, comments, and references. 
(1): Spice Library \citep{acto96}, (2): Spice Library and values in
units of volume mean radii $R_P$ for Earth: 6371.01 km, Jupiter: 69911 km,
Saturn 58232 km, Uranus: 25362 km,
 Neptune: 24624 km \citep{acto96},
(3):  \citet{hale12},  \citet{harn13},
(4): \citet{kive04}, 
(5): \citet{kive04}, \citet{khur98}, \citet{saur98}
(6): \citet{kive04}, 
(7): \citet{kive04}, \citet{stro02}, \citet{hart17a},   
(8): \citet{saur05}, 
(9): \citet{doug06}, \citet{krie09,krie11}, \citet{simo11},
\citet{saur07}, \citet{simo14},
(10): \citet{khur08}, \citet{simo09},
(11): \citet{saur05}, \citet{simo11a}, 
(12): \citet{khur08}, \citet{simo12}, \citet{khur17}
(13) \citet{neub06}, \citet{simo10}, \citet{arri11}, \citet{bert08}
(14) \citet{ness86}, \citet{brid86}, \citet{mcnu87}, 
(15) \citet{neub90}. 
}
 \label{tab:moons}
  \begin{tabular}{llrrrrrrrrrrr}
Planet&Satellite&Radius& Semi-Major& $B_0$ &$v_0$ & $n_p$ & $M_A$
    &$M_S$ &$\beta$&$\bar{\alpha}$ &Comments &Ref.\\
 & & km (1)& R$_P$ (2) & nT & km/s & cm$^{-3}$ & & & &Eq. (\ref{e:alpha_bar})\\ \hline
Earth &Moon& 1737.5      & 60.3& 10&100 &0.1 & 0.05&
                                                     $\sim$1&$<$0.01&-&20
                                                                        \%
                                                                        of
                                                                        orbit
                                                                        in
                                                                        magnetosphere
                                             &(3) \\
Jupiter&Io & 1821.3       &6.0&1720& 57&2500. & 0.31& 2.0 &
                                                   0.32
                   &0.9-0.95 & most powerful interaction
                                                                       &(4)\\
Jupiter & Europa & 1565 &9.6 & 370 &76 & 200 &0.47 &0.9 &0.32 &
    0.8 & induction in ocean & (5)\\
Jupiter & Ganymede& 2634 &15.3 & 64 & 139 &5.& 0.73 &0.5 & 2.4 & 0.2&
    intrinsic dynamo field, induction & (6)\\
Jupiter & Callisto &  2403    &26.9 & 4 & 192 &0.15 &2.8 &0.4 & 64
    &$>$0.99 &enhanced ionospheric conductances &(7)\\
Saturn & Mimas & 198.8   &3.2 & 722 &16 & 90 & 0.04 & 0.69&
                                                                   0.003& -
                                   & no in-situ observations & (8)\\
Saturn & Enceladus &252.3 &4.1 & 325 & 26&70.5 & 0.13 & 1.32
           &0.01 &$>$0.95& plumes, dusty plasma, hemisphere coup. &
                                                                    (9) \\
Saturn &Tethys & 536.3 &5.1 & 167& 34 & 30 & 0.21& 1.3 &
                                            0.03
                   & $<$0.03 &       wake effects              & (10)\\
Saturn & Dione &562.5          &6.5&75&40 & 13 & 0.37 &0.89 & 0.2&
                                                                   $\sim$0.03
                                   & weak interaction &(11) \\
Saturn & Rhea & 764.5     &9.1 & 25 &57 & 2 & 0.61 & 1.3 &
                                                    0.26& $<$0.15 &
                                                                    wake
                                                                    effects
                                             dominate&
                                                                       (12)\\
Saturn & Titan &2575.5 &21.0 & $\sim$5 & 120 &$\sim$0.1 &
                                                                    $\sim$1
    & $<$1 & $>$1 &1& highly variable plasma environ. & (13)\\
Uranus & Ariel &578.9             &7.5 & 80 &9
                                                      &0.15&0.002&0.02&0.01&-&
                                                                               no
                                                                               in-situ
                                             obs&(14)\\
Uranus & Umbriel &584.7    &10.5 & 30 &16&0.06&0.006&-& -&-&no in-situ
                                             obs&(14)\\
Uranus & Titania & 788.9          &17.2& 7&30&0.021&0.03&-& -&-& no
                                                                 in-situ
                                             obs&(164\\
Uranus & Oberon &761.4           &23.0 &3&42&0.016&0.09&-& -&-& no
                                                                in-situ
                                             obs& (14)\\
Neptune & Triton & 1352.6     &14.4 & 10&39 &-&0.1&-&-&-& no in-situ obs&(15)
  \end{tabular}
  \end{sidewaystable}
%
%

\subsection{Earth: Moon}
The Earth's moon orbits our planet with a semi-major axis of 60.3
$R_E$ (Earth radii). The Earth magnetopause on the sub-solar side is located at
roughly 10 $R_E$. Thus the moon is mostly located in the solar
wind and thus subject to a super-fast $M_f >1$ flow, but spends 20  \%
of its orbit in the terrestrial magnetosphere \citep{harn13}. During
its magnetospheric phase, the moon is subject to strongly
varying properties from nearly vaccum-like densities in the
magnetospheric lobes to the plasma sheet, in which number density and
heavy ion concentration can change on rapid time scales \citep{harn13}.  Since the moon
possesses only a very dilute exosphere, it acts mostly as an inert obstacle
to the flow but moon originating ions are still observed in the
Earth's magnetosphere. These studies date back to the sixties 
with more recent work published by, e.g., \citet{tana09} and \citet{hale12}. 

\subsection{Jupiter: Io, Europa, Ganymede, Callisto}
Jupiter possesses four large moons orbiting permanently within its magnetosphere:
Io, Europa, Ganymede and Callisto (Table
\ref{tab:moons}). As mentioned in Section \ref{s:intro}, Io's
interaction is historically the best studied moon-magnetosphere interaction likely because it is
the most powerful interaction. Io's interaction generates
magnetic field perturbations of $\sim700$ nT \citep{kive96}, it
it radiates away in each Alfv\'en wings a Poynting flux of $\sim$1
$\times$ 10$^{12}$ W \citep{saur13}.

Since all moons are in many aspects similar to
terrestrial type planets, there is a tremendous interest in using
observations of their space environment to understand the interior
structure of the moons.  For Io, \citet{khur11} argued that the magnetic
field measurements taken during  certain  Galileo spacecraft
flybys contain contributions which stem from induction in a 50 km
thick magma ocean very close to the surface. The
existence of such a surface near magma ocean  (although plausible from
considerations of Io's interior \citep{peal79}) 
has been
questioned based on Hubble Space Telescope observations of Io's
auroral spot oscillation and MHD modeling with appropriate
atmosphere models \citep{roth17,bloe18}. Both, Europa and Ganymede
exhibit induction signals from saline subsurface oceans
\citep{khur98,kive00,kive02}. In case of Ganymede, the induction
signal based on observations during a single flyby are however non-unique due to Ganymede's
internal dynamo magnetic field with its unknown quadrupole components.
This non-uniqueness could be solved and the ocean could be confirmed 
with HST observations of reduced
oscillation of Ganymede's  auroral ovals due to
induction in its saline subsurface ocean \citep{saur15}. Callisto was
also argued to possess a subsurface ocean based on induction signals
obtained during several Galileo spacecraft flybys
\citep{khur98,zimm00}. Recently the uniqueness of this interpretation
was questioned because of Callisto's highly conductive ionosphere
which can significantly contribute to Callisto's magnetic field
environment \citep{hart17a}.

Ganymede's interaction is unique because Ganymede's intrinsic dynamo
magnetic field causes a mini-magnetosphere within Ganymede's larger magnetosphere.
In contrast to the planetary magnetospheres for typical the solar wind
conditions, it is a magnetosphere without a bow shock, but with
Alfv\'en wings. Ganymede's magnetosphere is also particularly
interesting for reconnection studies because of the much more steady
setup compared to the interaction of the planetary magnetospheres with
the solar wind since Jupiter's and Ganymede's magnetic moments 
are approximately anti-parallel
\citep{neub98,ip02, paty04,jia08, popp18,zhan18,colli18,toth16}.

All Galilean moons possess dilute atmospheres \citep{stro05}. The momentum-exchange
of Jupiter's magnetospheric plasma with these atmospheres is therefore
the key reason for the moon-magnetosphere interaction in case of Io,
Europa and Callisto. Observations and
modeling of magnetic field and plasma perturbations \citep{bloe16,jia18,arno19}  
help to characterize 
the moons atmospheres \citep{hall95} and to search for 
plumes originally detected with the Hubble
Space Telescope 
\citep{roth14}. 
The mass loss from these atmospheres, in particular from Io and
Europa, are primary plasma sources for Jupiter's magnetosphere. The
most important mass loss is neutral sputtering with subsequent
ionization within the magnetosphere. Total loss rates are about
 10$^3$ kg s$^{-1}$ of SO$_2$ for Io and 50 kg s$^{-1}$ of O$_2$ for
 Europa \citep{bage11,dols08,saur98,mauk03,lagg03,dols16}.

The far-field interactions of the Galilean satellites are impressively
visible in form of satellite footprints 
at wavelength ranging from the infrared to the ultraviolet
(Figure
\ref{fig:Jupiter} and, e.g., \citet{conn93,pran96,clar96,inge98,clar02,bhat18}). The
footprints show a complex structure consisting of a main spot, a
leading spot, i.e., in front of the main spot, at footprint tail and
sometimes multiple spots \citep{bonf08,grod09}. The main spot is
caused by the primary Alfv\'en wing and its associated electron
particle acceleration. The leading spots result from trans-hemispheric
electron beams as part of the bi-directional electron acceleration
within the main wing \citep{bonf08}. The tails and the multiple spots
within the tails are generated by the reflected Alfv\'en waves at the
torus boundaries and other density gradients along the magnetospheric
field lines \citep{neub80,jaco07}. Depending on the strength of the
interaction at the moon (i.e., when the moon is in the magnetospheric
plasma sheet or outside), the reflection of the wave can be highly non-linear
where incident and reflection angle are different. The non-linear
interaction can smear out multiple spots in the tail
\citep{jaco07}. Very recently, observation with 
unprecedented spatial resolution in the infrared by the JUNO spacecraft
revealed detailed substructures of the multiple spots  \citep{mura18}.
These are
in appearance reminiscent of von Karman streets known from
hydrodynamics, but not in the underlying physics.

The physics of particle acceleration in the main wings and
  their tails is poorly constrained, but at the time of writing
  this chapter for the first time being probed with in-situ measurements by the Juno
  spacecraft.
Based on Juno measurements, \citet{szal18} found broadband bidirectional
electron beams in the high latitude wake region of Io consistent with
stochastic acceleration of reflected Alfv\'en waves \citep{hess10,bonf17a}.

\subsection{Saturn: Titan and icy satellites}

The moons of Saturn, which are permanently located within its magnetosphere are
Mimas, Enceladus, Tethys, Dione, Rhea. Titan can be located outside
the magnetosphere under exceptional solar wind conditions.
These bodies are all icy moons with Titan and Enceladus being arguably
the scientifically most exciting objects among Saturn's moons. 

Mimas, Tethys, Dione and Rhea are icy moons with thin exospheres.
These exospheres generate only a minor perturbation to the
magnetospheric plasma flow and magnetic field. However, these moons
act as absorbers of the upstream plasma and thus
generate empty wakes which are increasingly filled with plasma
further downstream driven by slow mode waves and accompanied by rarefaction
waves away from the wake. The associated movement of plasma and the
resultant magnetic field perturbation partially propagate away as
Alfv\'en waves contributing to the moons Alfv\'en wings
\citep{simo09,simo12,khur17}.

Enceladus provided a huge surprise to the Cassini Mission. Within four
rifts in its icy crust nicknamed 'tiger stripes', water vapor plumes
emerge. These plumes provide obstacles to the magnetospheric flow
and generate magnetic field perturbations, which were observed by the spacecraft
and successively led to the detection of the plumes with the other
Cassini instruments \citep{doug06,porc06}. 
These plumes are the
  major plasma source of Saturn's magnetosphere 
\citep{fles10,smit10}.
The plumes provide several
new interesting components of moon-magnetosphere interaction. Since the plumes are located below
the south pole, the northern Alfv\'en wing starts already in the
southern hemisphere. The northern wing is then partially blocked by
the absorbing insulating icy body, which leads to hemisphere coupling
currents and associated magnetic field discontinuities on field lines
tangent to the body \citep{saur07,simo14}. The plumes' gas contains a
significant amount of dust in form of mostly micrometer sized ice
particles, which are negatively charged, and thus render a complex
dusty plasma around Enceladus \citep{hill12}. The negatively charged dust
reverses the sign of the Hall conductivity within the plume \citep{simo11,krie11}.
Enceladus' plasma interaction
experiences time-variability due to the variability of the outgassing
from the plumes caused by tidal stresses along its eccentric orbit
around Saturn \citep{saur08,hedm13}. 

Titan is also an exceptional moon in many aspects. It possesses a nitrogen
rich atmosphere with an atmospheric surface pressure of 1.5 bar.  It
is thus the only moon where the surrounding plasma does not reach the
surface of the moon. It possesses an ionosphere and is subject to a
highly variable plasma and field field environment because Titan
orbits Saturn close to its magnetopause \citep{ryme09,simo10}. 
Due to the large plasma beta mostly as a results of the low magnetic field
strength at the orbit of Titan, the plasma is strongly draped around
Titan's atmosphere and ionosphere
\citep{ma06,modo08}.
Because of the very low plasma
velocities in Titan's ionosphere and the partly frozen-in magnetic
fields, time-variable external magnetic fields from previous times (up
to several hours) are still observable in its ionosphere. 
These fields are referred to as fossil fields
\citep{neub06,bert08}. For a comprehensive
discussion about Titan's interaction we refer the reader
to a separate chapter by C. Bertucci.

\subsection{Uranus: Icy moons}

Uranus harbors four moons with diameters larger than 1000 km within its
magnetosphere. They are Ariel, Umbriel,
Titania and Oberon. 
The only spacecraft to
perform in-situ observation of Uranus' magnetosphere was the Voyager
2 spacecraft, but it did not pass the moons at close enough distances to detect
signatures of moon-magnetosphere interaction. Voyager 2 found that
Uranus' magnetosphere,  
compared to those of Jupiter and Saturn, 
is
only
sparsely populated with plasma. The expected plasma and magnetic field values
near the moons are displayed in table \ref{tab:moons}. No atmospheres
or exospheres around the moons are observed. 
Thus the plasma interactions
 generated by the moons are expected to be very weak.

The large icy moons might harbor subsurface water
oceans \citep{huss06}. The magnetic moment of Uranus is tilted by approximately
$60^{\circ}$ with respect to its spin axis \citep{ness86}, which
will result in large amplitude time-periodic magnetic fields at the
locations of the moons. In case of electrically conductive oceans,
large dipole magnetic fields will be induced and be measurable
outside of the moons \citep{saur10}. These time-variable induced
dipole fields will also generate a weak far-field interaction in form of
Alfv\'en wings.

\subsection{Neptune: Triton}

In the Neptune system, Triton is the only moon with diameter larger
than 1000 km. Triton is a moon with a thin atmosphere and small scale
plumes out of methane. Thus the interaction of the magnetospheric
plasma with the atmosphere of Triton will generate an active,
i.e., $\bar{\alpha} > 0$ plasma interaction, which however has not been
observationally confirmed. Voyager 2 was so far the only spacecraft to
visit Neptune but without a sufficiently close encounter with Triton to
detect its moon-magnetosphere interaction.
Neptune's magnetic moment is inclined by 47$^{\circ}$ with respect to
its spin axis  \citep{ness89}. Triton in addition
possesses an inclination of 156.8$^{\circ}$. Therefore an observer in
the rest frame of Titan sees a highly time-variable magnetic field at
both the synodic rotation period of Neptune (14.46 h) and the orbital period of
Triton (5.88 d) \citep{saur10}. Any saline subsurface water ocean as
discussed in the literature \citep{huss06}, will generate easily
observable induction signals at close flybys.

\subsection{Extrasolar planets: Star-planet interaction}

The equivalent of moons in planetary magnetospheres in other stellar
systems are close-in extra solar planets interacting with their stellar
astrospheres.  If the close-in extra solar
planets are within the Alfv\'en radius  of the associated stellar
wind (defined by the location with
$M_A=1$), then the exoplanets are subject to sub-Alfv\'enic
interaction. Within that radius the interaction of the stellar wind
with the exoplanet can generate Alfv\'en wings which can reach the central star. This
counterpart to moon-magnetosphere interaction is called
electromagnetic 'star-planet
interaction'. Observational evidence comes, e.g., from correlation of stellar
Ca II emission with the orbital period of 
a 
close-in exoplanet
\citep{shko08}. Theoretical studies on star-planet interaction have
been performed numerically and analytically by 
\citet{cunt00,preu05,zark07,lanz08,saur13,stru15,turn18}.

\section{Outstanding questions}
\label{sec:questions}

Finally we discuss several open questions in the field of
moon-magnetosphere interaction:

The far-field interaction introduced in Section \ref{ss:farfield} is
observationally as well as theoretically the least understood part of
moon-magnetosphere interaction. It is in many aspects unclear how the Alfv\'en waves
evolve while traveling towards the planets. In particular, the
reflection and filamentation processes which are expected to take
place are of key interest \citep{chus05,hess10,hess13a}.
 Also the acceleration processes where
the Alfv\'en wave energy is converted into accelerated electrons and ions, which lead
to the satellite footprint emission in the planets' atmospheres has been barely
investigated. Here relativistic effects also play a role in parts
of the magnetospheres with $v_A$ from (\ref{e:vA}) assuming values
larger then the speed of light and with electron energies larger than a few
hundred keV.
Further studies will hopefully also shed light into understanding
the enigmatic tail features of the Io footprints recently observed by
the JUNO spacecraft \citep{mura18}.
The coupling of the
  Alfv\'en wings to the planets' ionospheres is also a very poorly
  studied subject. Models for the coupling between the wings and
  the ionospheres will require the inclusion of the steep density 
  gradients and the consideration of anisotropic ionospheric
  conductivities.

Even though the local interaction introduced in Section \ref{ss:local}
is much better understood than  the far-field interaction, the
non-linear interaction of the various processes, such as plasma
interaction in the atmospheres, induction in the interiors and ionospheres, and
atmosphere-plasma feedback are only poorly investigated. The later
points address how the plasma interaction contributes to the
generation, loss and reshaping of the dilute moon-atmospheres.

Ultimately, it will be necessary to combine the near-field, the
  far-field and the coupling to the planets' ionosphere for obtaining
  a 'global' understanding of moon-magnetosphere interaction. This
  includes an understanding of the linear momentum, angular momentum
  and energy fluxes between the moons and their planets.

The moon-magnetosphere interaction of the moons of  Uranus and Neptune
are observationally only weakly constrained and therefore new space missions
to these two planets are highly desirable \citep{arri14,chri12}.

A better cross-comparison of moon-magnetosphere interaction and star-planet
interaction in extrasolar systems would also be highly desirable as
the similarities and differences taking place in these systems are not
sufficiently investigated. Efforts in this area would additional help
to bridge the space physics community of the solar system with the
astronomical community
and will contribute to unify planetary sciences across their borders.

\acknowledgments
The author is grateful to Fritz M. Neubauer for 
introducing him to the field of moon-magnetosphere interaction and the 
valuable discussions over
decades. 
The interaction with his students and postdocs and the discussion with
research colleagues over
the years is also greatly appreciated.


%
%





\end{document}